\documentclass[twocolumn,secnumarabic,amssymb, nobibnotes, aps, prd]{revtex4-1}

\newcommand{\rv}{{\bm r}}
\newcommand{\xv}{{\bm x}}

\newcommand{\lambdav}{\bm\lambda}

\renewcommand{\v}{\bm}
\usepackage{amsmath,amssymb}
\usepackage{graphicx} 
\usepackage{longtable}

\begin{document}


\title{\textbf{Hard sphere packings within cylinders}}


\author{Lin Fu,\textit{$^{a}$} William Steinhardt,\textit{$^{b}$} Hao Zhao,\textit{$^{a}$} Joshua E.~S.~Socolar,\textit{$^{b}$} and Patrick Charbonneau$^{\ast}$\textit{$^{ab}$}}
\affiliation{\textit{$^{a}$~Duke University, Department of Chemistry, Durham, NC 27708, USA. E-mail: patrick.charbonneau@duke.edu}, \textit{$^{b}$~Duke University, Department of Physics, Durham, NC 27708, USA.}}


\date{\today}

\begin{abstract}
The packing of hard spheres (HS) of diameter $\sigma$ in a cylinder has been used to model experimental systems, such as fullerenes in nanotubes and colloidal wire assembly. Finding the densest packings of HS under this type of confinement, however, grows increasingly complex with the cylinder diameter, $D$. Little is thus known about the densest achievable packings for $D>2.873\sigma$.  In this work, we extend the identification of the packings up to $D=4.00\sigma$ by adapting Torquato-Jiao's adaptive-shrinking-cell formulation and sequential-linear-programming (SLP) technique.  We identify 17 new structures, almost all of them chiral. Beyond $D\approx2.85\sigma$, most of the structures consist of an outer shell and an inner core that compete for being close packed.  In some cases, the shell adopts its own maximum density configuration, and the stacking of core spheres within it is quasiperiodic.  In other cases, an interplay between the two components is observed, which may result in simple periodic structures. In yet other cases, the very distinction between core and shell vanishes, resulting in more exotic packing geometries, including some that are three-dimensional extensions of structures obtained from packing hard disks in a circle.
\end{abstract}


\maketitle

\section{Introduction}
Packing problems, i.e., identifying optimal configurations of a set of hard geometric objects in space without overlap, are NP-hard optimization problems~\cite{hifi2009review}. Of these problems, packings of hard spheres (HS) have been the most extensively investigated because of their dual physical generality and mathematical elegance. For instance, in condensed matter hard sphere packings help explain both the ordering of binary alloys~\cite{hopkins2011,filion2009pre,kummerfeld2008} and the persistence of disorder in metallic glass formers~\cite{zhang2014}. Mathematical demonstrations of the optimality of these packings are also prestigious challenges to surmount. In two dimensions, the triangular lattice has long been known to be the densest packing of disks~\cite{aste2008}, but the proof of the Kepler conjecture for hard spheres in three dimensions was only recently obtained~\cite{hales2005proof} (with much fanfare~\cite{aste2008}), and in higher dimensions demonstrations are an absolute scarcity~\cite{conway1999,Cohn2003}. 

Although packing objects in infinite spaces can be a reasonable model for describing the structure of bulk materials, in confined systems boundaries play a substantial role. The consideration of packings in bounded spaces is also motivated by the wide variety of situations in which objects are stored in containers of different shapes~\cite{stoyan1982,fraser1994,yeung2005hybrid}. Many variants of this problem have thus been considered by mathematicians and physicists, including disks in a circle~\cite{graham1998circle,fodor2003circle,hopkins2010pre}, a rectangle or a strip~\cite{hifi2013strip,yaskov2004strip} in two dimensions, and spheres in a sphere, a cube~\cite{goldberg1971cube,gensane2004cube} or a parallelepiped~\cite{yaskov2003parallelepiped} in three dimensions. Irrespective of the dimensionality of the packed objects, however, the above packing problems are all quasi-zero-dimensional, because extrapolating to the small system limit gives a simple point. Packing objects between planes, i.e., in quasi-two-dimensional confinement, has also been intensely studied~\cite{pieranski1983,schmidt1996,fortini2006}, notably in order to understand the transition from two-step to first-order melting as the thickness is increased from a monolayer to a bulk system. 

The intermediate case of quasi-one-dimensional confinement models a number of experiments, including packing fullerenes in nanotubes~\cite{zettle2003science,briggs2004prl}, colloidal wire formation~\cite{meseguer2008advmat} and nanoparticle self-assembly in cylindrical domains~\cite{nandan2014angewandte,zhu2014macro}, yet has received substantially less attention than other confinement types. Existing studies nonetheless suggest that these systems can display a variety of remarkable structural features. 
The first efforts by \citeauthor{pickett2000PRL} to systematically identify maximally dense packings of HS of diameter $\sigma$ in cylinders of diameters $D$ revealed the emergence of spontaneous helical chirality~\cite{pickett2000PRL}; and \citeauthor{mughal2012PRE}'s systematic extension of this work from $D=2.155\sigma$ to $D=2.873\sigma$ showed how plane disclinations allow helical structures to continuously transform into one another~\cite{mughal2012PRE}. 

Because the largest cylinders studied remain quite far from the bulk limit, a rich set of structural features is to be expected in wider cylinders as well. Many of the existing tools for finding optimal packings are, however, insufficient in this regime. Earlier analyses relied on simulated annealing~\cite{pickett2000PRL,mughal2012PRE}, whose computational efficiency is too limited for finding packings with a separate inner core and outer shell. Algorithms based on sequential deposition are much more expedient~\cite{chan2011PRE}, but the final structure they generate depends too sensitively on the choice of underlying template for them to be of broad applicability. Although genetic algorithm schemes (akin to that used in Ref.~\cite{filion2009pre}) can search configuration space more broadly, the growing structural complexity of the packings with $D$ necessitates unit cells that are too large for the approach to remain computationally tractable.

In this work, we improve the computational capability of finding dense HS packings cylinders with $D>2.873\sigma$ by instead using an adaptive-shrinking cell and a sequential-linear-programming (SLP) technique~\cite{torquato2010PRE}. This approach confirms earlier results for $D\leq2.86\sigma$ and extends our knowledge of packings up to $D=4.00\sigma$. Interestingly, many of the new structures appear to be quasiperiodic and another of the packings presents fairly exotic geometries. In Section~\ref{sec:method}, we describe the SLP computational method, in Section~\ref{sec:compresults} we present the SLP results for different ranges of cylinder diameters, and in Section~\ref{sec:analytical} we specifically analyze the quasiperiodic structures observed in the range $2.988\sigma\leq D\leq 3.42\sigma$ using a sinking algorithm developed for this problem.
\section{Sequential Linear Programming Method}
\label{sec:method}
In order to identify HS packings in cylinders, we adapt the SLP method of Torquato and Jiao~\cite{torquato2010PRE} to this geometry. For convenience, we describe configurations using cylindrical coordinates with z, r, and $\theta$ representing the axial, radial and angular components, respectively. For our search procedure, we consider a fixed number of spheres in a finite cylinder with periodic boundary conditions that match the top of the cylinder to the bottom with a twist.  In other words, the entire volume of the cylinder is initially taken as a unit cell with a one-dimensional periodicity.  The infinite structure consists of spheres centered at
\begin{equation}
\rv_{ij}=\rv_{i}+n_{j}\lambdav~,
\end{equation}
where $\rv_{i}$ ($i=1,2,...$) are the particle positions within a unit cell, $n_{j}\in\mathbb{Z}$, and $\lambdav=(\lambda_{\mathrm{r}},\lambda_{\mathrm{\theta}} ,\lambda_{\mathrm{z}})$ is the lattice vector.  For our system, $\lambda_{\mathrm{z}}$ is the height of the unit cell, $\lambda_{\mathrm{\theta}}$ is the twist angle, and $\lambda_{\mathrm{r}}=0$. 
The volume of a unit cell is thus
\begin{equation}
v_{\mathrm{u}}=\pi (D/2)^{2}\lambda_{\mathrm{z}}~,
\end{equation}
and the lattice packing fraction can be expressed as
\begin{equation}
\eta = \frac{Nv_{\mathrm{s}}}{v_{\mathrm{u}}}~,
\end{equation}
where $N$ is the number of particles in a unit cell and $v_{\mathrm{s}}$ is the volume of a sphere of diameter $\sigma$.

A  each optimization step, we allow the $N$ particles in the unit cell to move as well as changes to the unit cell height, $\lambda_{\mathrm{z}}$, and twist angle, $\lambda_{\mathrm{\theta}}$. Denoting the particle displacements $\Delta\rv$, the matrix specifying changes to the unit cell $\v{\epsilon}$, the new particle positions $\rv^{\mathrm{n}}$, and the new lattice vector $\lambdav^{\mathrm{n}}$, we have
\begin{eqnarray}
\lambdav^{\mathrm{n}}&=&(\v{I}+\v{\epsilon})\lambdav\\
\v{\epsilon}&=&\begin{bmatrix} 0 & 0 & 0 \\ 0 & \epsilon_{\mathrm{\theta}} & 0 \\ 0 & 0 & \epsilon_{\mathrm{z}} \end{bmatrix}\\
\rv^{\mathrm{n}}&=&(\ r_{\mathrm{r}}+\Delta r_{\mathrm{r}},r_{\mathrm{\theta}}+\Delta r_{\mathrm{\theta}},(1+\epsilon_{\mathrm{z}})(r_{\mathrm{z}}+\Delta r_{\mathrm{z}})\ )~.
\end{eqnarray}
Note that the twist angle is not continuously shearing the particles within the unit cell, but is only a property of the boundary conditions. It thus only appears in $\rv^{\mathrm{n}}$ for boundary particles. 
In order to find the maximum packing density we must solve the following problem:
\[\mathrm{minimize}~v_{\mathrm{u}}\]
subject to 
\begin{eqnarray}
r^{\mathrm{n}}_{mn}&\ge&\bar{D}_{mn} ~,~\forall~mn~\mathrm{neighbor~pairs},\\
r_{i\mathrm{r}}+\Delta r_{\mathrm{r}}+R_{i}&\le&D/2,~\forall~i=(1,2,...,N),
\end{eqnarray}
where $\bar{D}_{mn}=(D_{m}+D_{n})/2$ and $R_{i}$ is the radius of particle $i$. The first condition corresponds to the hard-sphere constraint and the second to the hard-wall constraint. Because during the optimization $D$ is fixed, minimizing $v_{\mathrm{u}}$ is equivalent to minimizing $\lambda_{\mathrm{z}}$. The packing problem then becomes a standard constrained optimization problem, and the constraints can be linearized (by Taylor expansion), allowing the use of linear-programming solvers~\cite{torquato2010PRE, glpk}.
The optimization problem at each step then becomes
\[\mathrm{minimize}:~\epsilon_{\mathrm{z}},\] 
subject to
\begin{align}
r^{\mathrm{n}}_{mn}&\ge\bar{D}_{mn},\forall~mn~\mathrm{neighbor~pairs}\\
D/2&\ge r_{i\mathrm{r}}+\Delta r_{\mathrm{r}}+R_{i},~\forall~i\\
\Delta r^{\mathrm{lower}}&\le|\Delta\rv_{i}|\le\Delta r^{\mathrm{upper}},~\forall~i\\
\epsilon^{\mathrm{lower}}&\le|\v{\epsilon}|\le\epsilon^{\mathrm{upper}},
\end{align}
where
the superscripts refer to the upper and lower bounds for particle displacements and volume changes. Although the solution of this problem can be obtained by linear programming, displacements and unit cell adjustments at each optimization step must remain relatively small because of the linearization. The overall packing optimization must thus be done sequentially, meaning that the optimal solution for a given step is used as input for the subsequent one, until convergence is achieved. Operationally, we accept a solution as having converged when the difference between two iterations $|\Delta v_{\mathrm{u}}|<v_{\mathrm{tol}}$, where $v_\mathrm{tol}=10^{-6}$. Note that because this criterion is independent of the unit cell size, the final unit cell volume is determined less precisely for larger unit cells, but this effect is negligible on the scale of the figures and other numerical results reported here.

\section{SLP Results}
\label{sec:compresults}

\begin{figure}[h]
\centering
  \includegraphics[height=7cm]{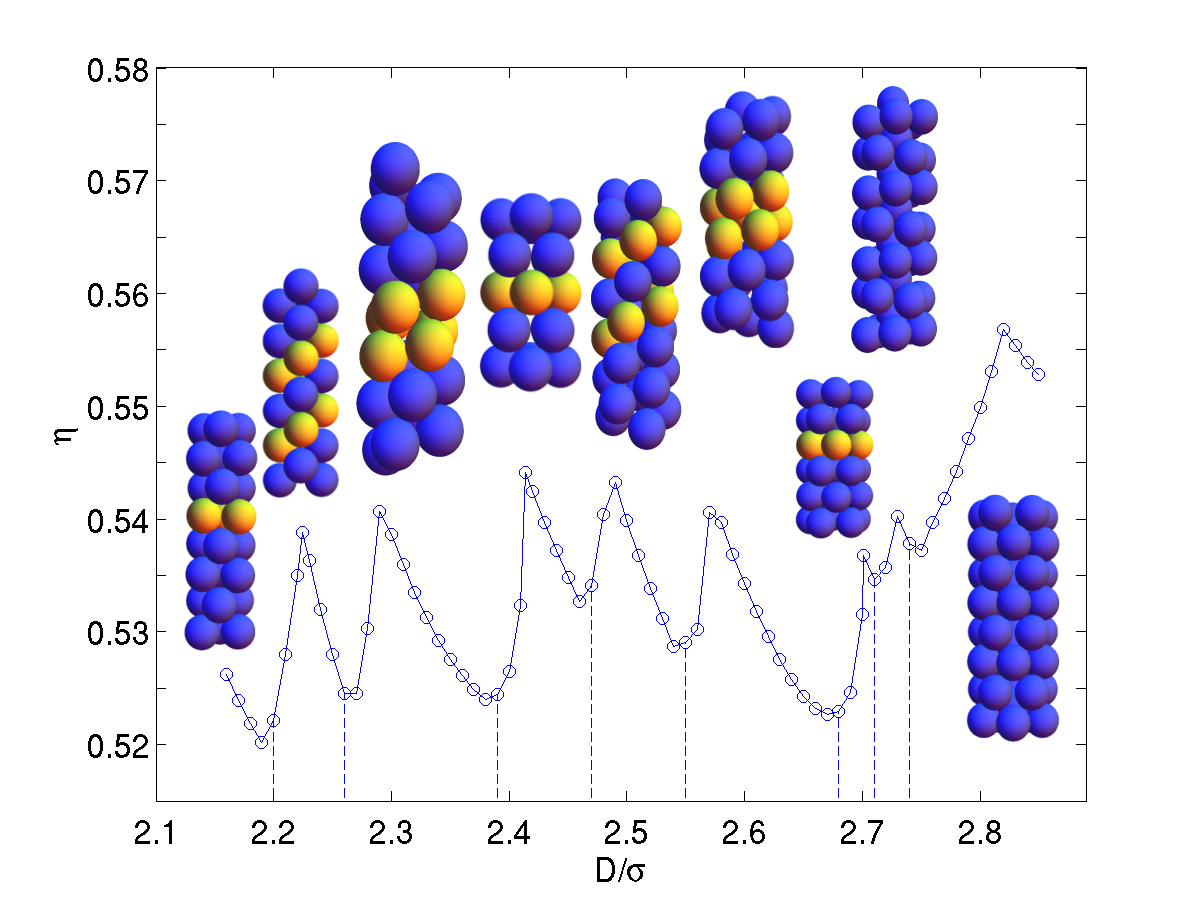}
  \caption{Packings for $2.16\sigma\leq D\leq2.86\sigma$. Configurations are depicted at the density maxima. Yellow particles are part of a same helix. In this regime, the results are in complete agreement with those of Ref.~\cite{mughal2011prl}. Figure~\ref{fig:285400} presents results for $D\geq 2.86\sigma$.}
    \label{fig:216285}
\end{figure}

Using the SLP method, we identified candidate structures for HS packings from $D=2.16\sigma$ to $D=4.00\sigma$. For $D\leq2.862$, we reproduced previously reported results~\cite{mughal2012PRE} (Fig.~\ref{fig:216285}), but we obtained denser structures than \citeauthor{mughal2012PRE} for $2.862\sigma<D<2.873\sigma$ (Fig.~\ref{fig:285400} inset). This discrepancy is likely due to the difference in system sizes between the two studies. The packings obtained for this regime in Ref.~\cite{mughal2012PRE} had unit cells with either $N=7$ or 15 particles, while our search extended up to $N=150$, and the densest structures had $50\leq N\leq85$. The origin of this strong system size dependence likely lies in the aperiodicity or the complex periodicity of the packings. (We come back to this point in Section~\ref{sec:286300}.) For $D\geq2.873\sigma$, no systematic studies had previously been undertaken, and only a few structures had been proposed~\cite{huang2009JCP}. SLP identifies 17 distinct structures and their deformations over $2.873\sigma\leq D\leq4.00\sigma$ (Fig.~\ref{fig:285400}). The packings depicted in Figure~\ref{fig:285400} are local maxima in $\eta(D)$. 

\begin{figure}[h]
\centering
  \includegraphics[height=5.5cm]{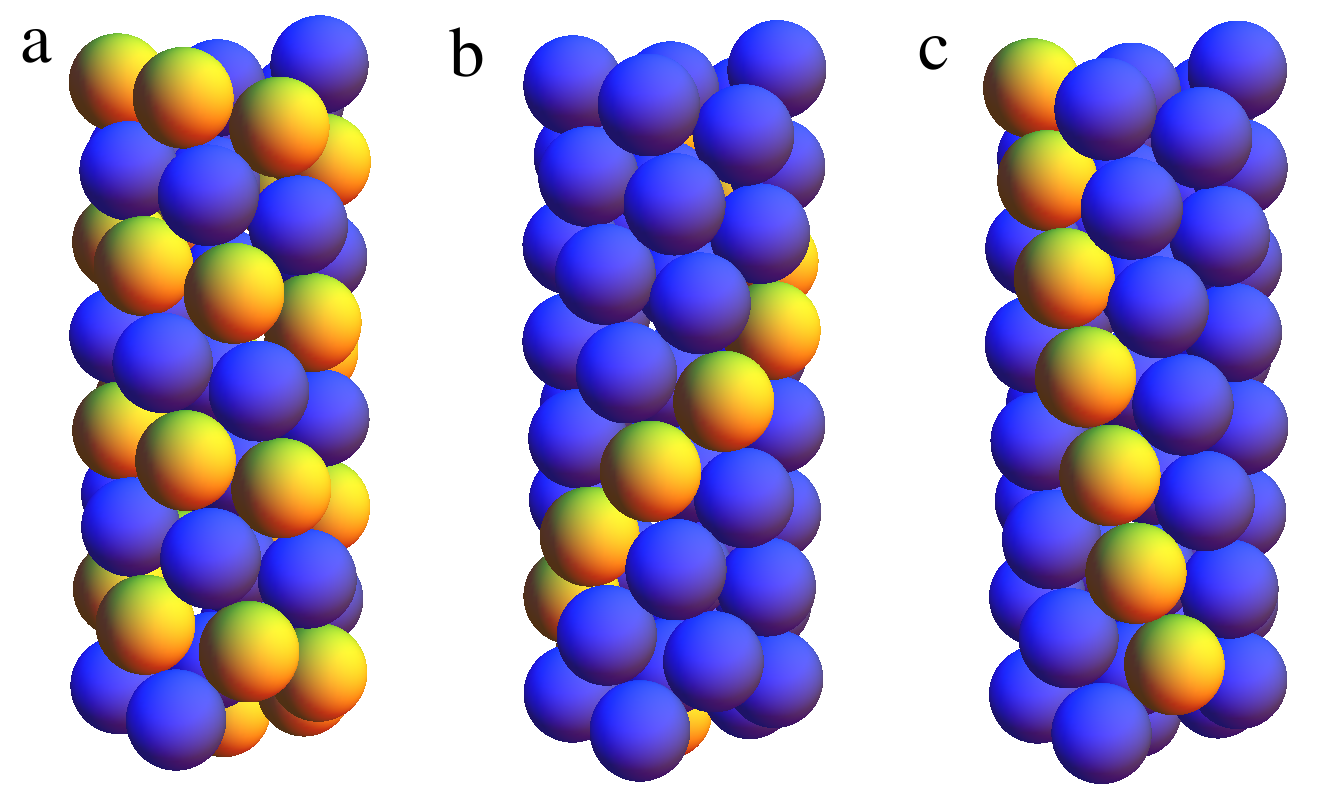}
  \caption{There exists three different ways to define a helix, and thus six possible line-slip structures. Yellow particles are part of a same helix. In the text, we select the convention depicted in (a).}
  \label{fig:helixdef}
\end{figure}

Most of the structures in this regime have two well-defined layers: an inner core and an outer shell. Because many of the outer shells are optimal packings of disks on the inner surface of the cylinder, they can be described using the phyllotactic notation for helices, $(l,m,n)$, with $l=m+n$, where $l$, $m$ and $n$ are the number of helices, using the three possible helix definitions (Fig.~\ref{fig:helixdef})~\cite{mughal2011prl}. The parameters defining some of these helical outer shells are listed in Table~\ref{tbl:outer}. For simplicity, we denote below structures with the helix whose height difference, $\Delta z$, between two successive particles within that helix is minimal, and $\Delta\theta$ is thus the angular coordinate difference between two successive particles within that helix (Fig.~\ref{fig:helixdef}a). Based on these definitions, we note that $\lambda_z = (N_s/l) \Delta z$, $\lambda_{\theta} =  \mod_{2\pi}\left((N_s/l) \Delta\theta\right)$, where $N_{\mathrm{s}}$ is the number of shell particles in the unit cell.

Intermediate structures can be obtained by continuously transforming the local density maxima. Some are uniform radial expansions (or compressions) of these structures, accompanied with a compression (or expansion) in $\mathrm{z}$, while other structures undergo a line-slip, which is a slip between two helices, keeping the relative position of the other helices constant~\cite{mughal2012PRE}. Because a helix can be defined in three different ways (Fig.~\ref{fig:helixdef}), each maximal density structure presents up to six corresponding line-slip possibilities (two directions for each type of slip), although symmetry can reduce this number. 

In the following subsections we present an overview of different $D$ regimes over which the packings we obtain share a number of structural features.

\begin{table}[h]
\small
  \caption{\ Structural parameters and properties of close-packed outer shells for different cylinder diameters. Quantities are rounded to the last digit}
  \label{tbl:outer}
  \begin{tabular}{@{\extracolsep{\fill}}c|c|c|c|c|c}
    \hline
    Notation & $D/\sigma$ & $\Delta\theta$ & $\Delta z/\sigma$ & Chirality & Number of helices\\
    \hline
    (6,5,1) & 2.8652 & 1.1167 & 0.1538 & chiral & 1\\
    (6,6,0) & 3.0000 & 1.0472 & 0.0000 & achiral & /\\
    (7,4,3) & 3.0038 & 0.9365 & 0.4268 & chiral & 3\\
    (7,5,2) & 3.0623 & 0.9697 & 0.2759 & chiral & 2\\
    (7,6,1) & 3.1664 & 0.9507 & 0.1309 & chiral & 1\\
    (8,4,4) & 3.2630 & 0.7854 & 0.5000 & achiral & 4\\
    (8,5,3) & 3.2888 & 0.8357 & 0.3706 & chiral & 3\\
    (7,7,0) & 3.3048 & 0.8976 & 0.0000 & achiral & /\\
    (8,6,2) & 3.3615 & 0.8475 & 0.2391 & chiral & 2\\
    (8,7,1) & 	3.4720 & 0.8272 & 0.1139 & chiral & 1\\
    (9,5,4) & 3.5377 & 0.7220 & 0.4434 & chiral & 4\\
    (9,6,3) & 3.5818 & 0.7496 & 0.3267 & chiral & 3\\
    (8,8,0) & 3.6131 & 0.7854 & 0.0000 & achiral & /\\
    (9,7,2) & 3.6648 & 0.7512 & 0.2109 & chiral & 2\\
    (9,8,1) & 3.7805 & 0.7318 & 0.1008 & chiral & 1\\
    (10,5,5) & 3.8025 & 0.6283 & 0.5000 & achiral & 5\\
    (10,6,4) & 3.8223 & 0.6624 & 0.3971 & chiral & 4\\
    (10,7,3) & 3.8800 & 0.6771 & 0.2917 & chiral & 3\\
    (9,9,0) & 3.9238 & 0.6981 & 0.0000 & achiral & /\\
    (10,8,2) & 3.9711 & 0.6738 & 0.1883 & chiral & 2\\
    \hline
  \end{tabular}
\end{table}

\subsection{Structures for $D<2.86\sigma$}

In this regime, all the structures are periodic and have a simple mathematical description. Most of them are simple helices. The last two structures, however, are non-helical and contain an inner core (Fig.~\ref{fig:216285}). The densest structure for $2.71486\sigma\le D\le2.74804\sigma$ has D$_{5}$ symmetry with a close-packed inner core, and  can be constructed as a packing of spindles of nearly regular tetrahedra. The optimal structure for $2.74804\sigma\le D\le 2.8481\sigma$ has instead a unit cell of 11 particles -- an inner particle sandwiched between two staggered five-particle rings -- that is reminiscent of a stacking of ferrocene molecules. Note, however, that neither the inner core nor the outer shell of this last structure are close-packed.

\subsection{Structures for $2.86\sigma\le D<2.988\sigma$}
\label{sec:286300}
\begin{figure*}[t]
\centering
  \includegraphics[width=2.0\columnwidth]{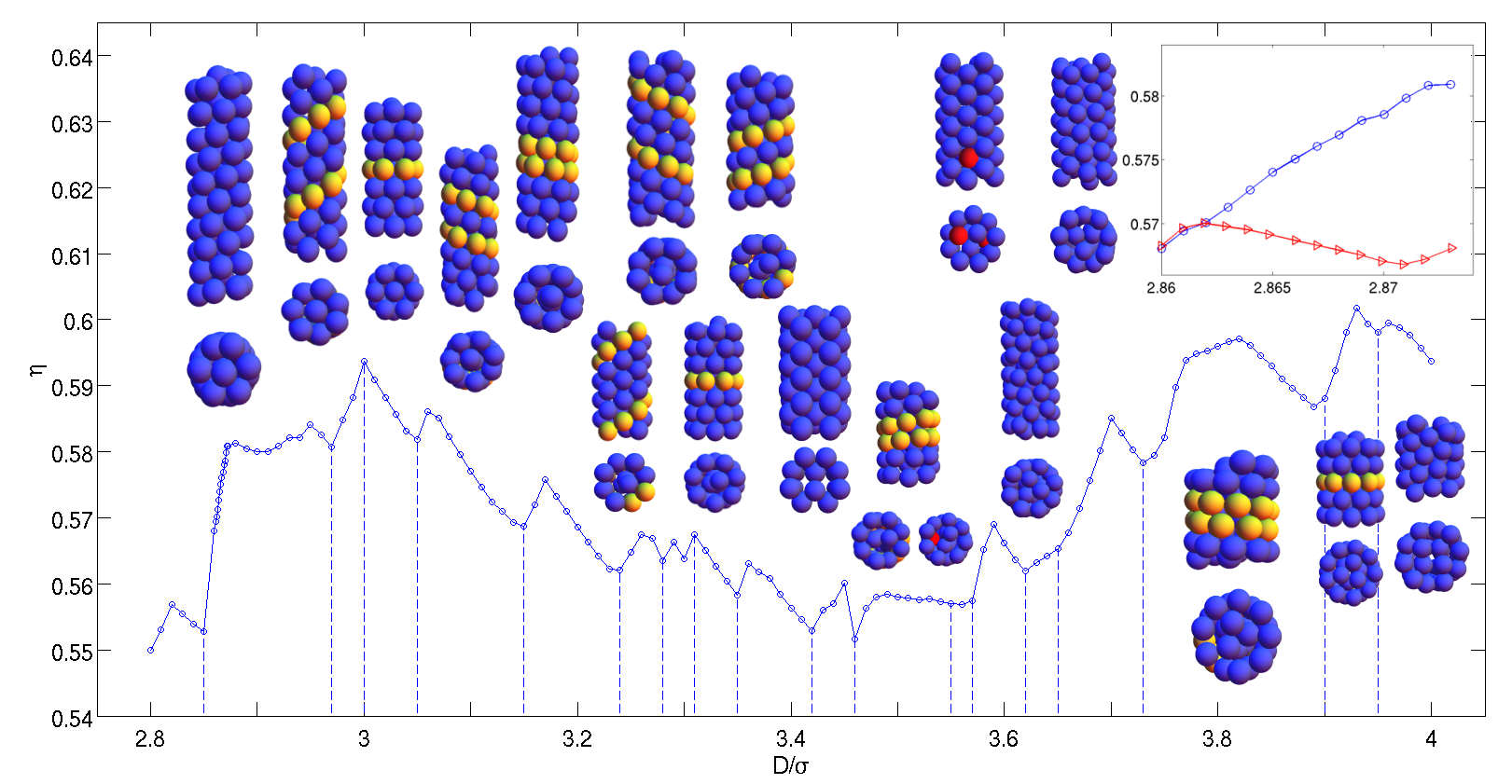}
  \caption{Packings for $D=2.85 - 4.00\sigma$. Configurations are depicted at the density maxima. Yellow particles are part of a same helix or its line-slip structure, as described by the phyllotactic notation, and red particles are hoppers (see text for details). Note that some configurations do not have a well-defined helical structure. The inset shows the difference between \citeauthor{mughal2012PRE}'s (red triangles) and the current (blue circles) results for $2.86\sigma<D<2.875\sigma$.}
  \label{fig:285400}
\end{figure*}

For $D\ge2.86\sigma$, packings do not have simple analytical descriptions. The competition between the inner core and the outer shell becomes more complex, because the two layers have different packing requirements, and neither of them systematically wins. For $2.86\sigma\le D\le 2.988\sigma$, the inner core dominates. To see why, note that a core particle can only fit into a shell formed by a horizontal layer of 6 spheres when $D\ge3\sigma$. For $D<3\sigma$, a core sphere can only fit if these 6 spheres form a helix around it. Every core particle must thus be at the center of a six-particle helix, which limits its freedom to move within the inner core. For instance, for the close-packed (6,5,1) outer shell, for which $D=2.8652\sigma$, the spacing between two turns of the six-particle helix that forms the outer shell is $6\Delta z=0.9228\sigma<\sigma$. If the outer shell were close-packed, then three six-particle helices would only accommodate a single inner particle, leaving large gaps between inner particles. A denser packing is instead obtained by deforming the outer shell in order to accommodate a close-packed inner core.

As $D$ approaches $3\sigma$, core particles become increasingly free to move. For $2.97\sigma<D<3.00\sigma$,  the densest possible outer shell is a line-slip structure of (7,4,3). The perfect (7,4,3) outer shell at $D=3.0038\sigma$ has an inner core spacing of $\frac{7}{3}\Delta z=0.9959(1)\sigma$, which is barely smaller than a particle diameter. Hence, for $2.97\sigma<D<2.988\sigma$, although the overall structure remains dominated by the inner core, the outer shell is barely different from a close-packed line-slip structure of (7,4,3).

\subsection{Structures for $2.988\sigma\le D\le3.42\sigma$}
\label{sec:3-342}

For $2.988\sigma\le D$, the inner core is sufficiently large to allow core particles to move freely within an outer shell, thus greatly reducing their constraint on the outer shell. Note that the lower end of this range is smaller than 3$\sigma$ because the densest outer shell for $2.988\sigma\le D<3.000\sigma$ is a line-slip structure of (7,4,3), for which no six outer particles are ever in the same plane. They can thus wrap an inner core without difficulty. Optimal packings from that point on and up to $D=3.42\sigma$ are found to almost always form the densest possible shell packing, independently of the core particles. The local density maxima in Figure~\ref{fig:285400} for this regime indeed all correspond to the diameters of close-packed outer shells (Table~\ref{tbl:outer}).

\begin{figure}[h]
\centering
  \includegraphics[height=6cm]{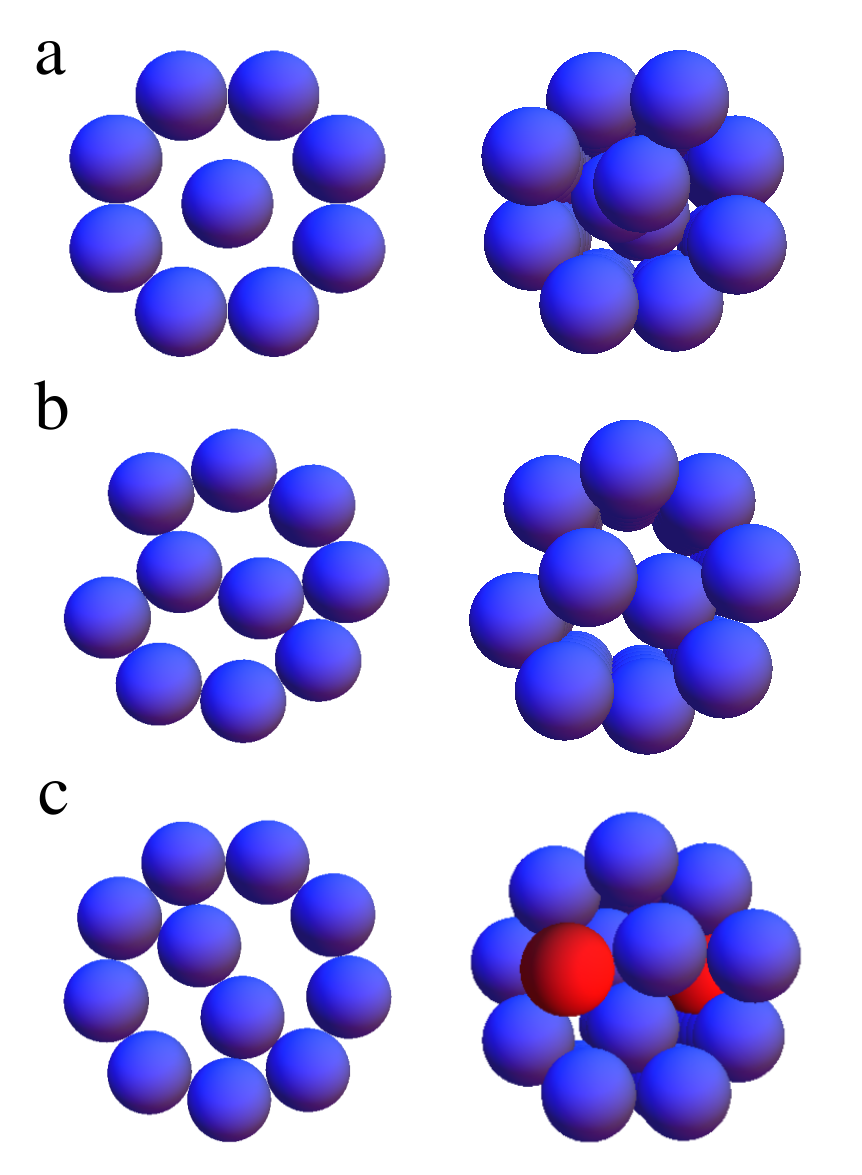}
  \caption{Comparison between disks in a circle and the top view of spheres in a cylinder at (a) $D=3.613\sigma$~\cite{pirl1969} (circle) and $D=3.25\sigma$ (cylinder), (b) $D=3.813\sigma$~\cite{pirl1969} (circle) and 3.43$\sigma$ (cylinder), and (c) $D=3.924\sigma$~\cite{melissen1994} (circle) and $3.58\sigma$ (cylinder). Red particles are hoppers. The cylinder outer shells consist of straight columns and only the top layer of particles is visible, so the resulting packings look similar to those of disks in a circle. Note that because the height of neighboring columns is shifted by $0.5\sigma$, the cylinder diameter is smaller than that of the circle and projecting particles onto the cylinder base reveals overlaps.}
  \label{fig:circlevscylinder}
\end{figure}

Out of the sequence, the structure with a (8,4,4) outer shell is particularly noteworthy. 
As for all outer shells with $m=n=\frac{1}{2}l$, this helical structure is achiral -- two of the three possible helical directions are equivalent, and $\Delta z=\sigma/2$. As a result, the outer shell consists of straight columns when viewed from the top of the cylinder. The top view of (8,4,4) outer shell and its core is thus very similar to the packing of disks in a circle (Fig.~\ref{fig:circlevscylinder}a). Two other structures are found to have this property (Fig.~\ref{fig:circlevscylinder}b and c), but the structure with the (8,4,4) outer shell is the only one that is close-packed. Note that a similar phenomenon would likely be observed for a structure with a (10,5,5) outer shell were it to be a density maximum (which it is not).

The case $D=3.00\sigma$ is also remarkable. The outer shell is then a close-packed structure with staggered six-particle rings, i.e., (6,6,0). The spacing between two rings is $\Delta_\perp=\sigma\sqrt{\sqrt{3}-1}\doteq0.8556\sigma$. 
Although core particles placed between the planes of the rings can shift off the cylinder axis, they cannot shift enough to allow a periodic packing of the core with no gaps between successive core spheres. This phenomenon illustrates the difficulty of searching for close-packed structures in this regime. Optimal structures may be quasiperiodic and thus not correspond to any finite $\lambda_{\mathrm{z}}$. Our numerical approach then at best provides a periodic approximant of the optimal structure. In order to consider this issue more carefully, we present an alternate algorithm for studying this regime in Section~\ref{sec:analytical}. 

\subsection{Structures for $D>3.42\sigma$}
For $D>3.42\sigma$, many of the close-packed outer shells are not observed. Instead of remaining disordered or quasiperiodic, the inner core then forms nearly ordered structure, which imposes many defects on the outer shell. For some of the packings, the defects are so large that they enable the two shells to interpenetrate. The packing structures in this regime are thus not clearly dominated by any one of the two layers, hence neither of the two shells is typically close-packed. For instance, of $l$-particle staggered ring structures, $(l,l,0)$, only $l=6$, 7 and 9 are observed. For $l=6$ and 7, the inner core is so small that only a lightly zig-zagging chain of particles fits within it; for $l=9$, the inner core is large enough that a staggered three-particle ring structure fits. For $l=8$, however, the zig-zagging structure is not very dense, and a two-particle flat pair does not fit. The packing structure thus ends up having a completely different organization: a dense triple helix inner core and an tortuous outer helical shell of eight particles.

The competition between the two shells does not only result in defective compromises, but also yields two novel types of structures. First, some structures are analogous to three-dimensional extensions of packing of hard disks in a circle (Fig.~\ref{fig:circlevscylinder}b and c).  Although the roughly straight columns formed by these structures gives their top view a two-dimensional feel, they are not simple stacks of these packings. As can be seen in Figure~\ref{fig:circlevscylinder}, projections of particles onto the cylinder base reveals overlaps. The outer shell is an (imperfect) triangular lattice rather than a square lattice, and the outer rings are not flat but form zig-zags. As a result the same three-dimensional version fits in a cylinder of a smaller diameter than that of the two-dimensional circle.
Second, some structures cannot be neatly divided into shells. For instance, for $3.55\sigma\le D\le3.61\sigma$, although both layers are dense the gap between them is sufficiently large to allow outer particles to hop back and forth between the two shells, keeping $\eta$ unchanged (Fig.~\ref{fig:circlevscylinder}c and Fig.~\ref{fig:285400}).

Table~\ref{tbl:outer} indicates that four ten-fold ($l=10$) outer shells could potentially be observed for $D\le4.00\sigma$. Yet only one appears in the phase sequence, as the last structure. The other three structures are missing, because for $3.62\sigma\le D\le3.94\sigma$, the inner core is just large enough to accommodate a triple helix, and thus a nine-fold helical outer shell provides a better periodicity than a ten-fold one to accommodate this inner shell.

As discussed in Sec.~\ref{sec:286300}, for $3.00\sigma\le D\le3.42\sigma$, the incommensurability of the two shells may result in structures that are not periodic. The densest structures thus cannot be obtained using a finite periodic unit cell, which makes the numerical search for packings by SLP extremely challenging (if not impossible) for some regions, e.g. $3.00\sigma\le D\le3.05\sigma$. By contrast, for $D>3.42\sigma$ the strong interaction between the two shells forces the two to share a same periodicity in some ranges of $D$ (see Supplementary information). The packings in those regions are thus periodic, which reduces the computational difficulty of identifying these structures. The vastness of the configurational space to sample at this point, however, reduces the confidence with which truly optimal packings are then identified.

\section{Sinking algorithm for $2.988\sigma\le D\le3.42\sigma$}
\label{sec:analytical}

SLP results suggest that for $2.988\sigma\le D\le3.42\sigma$ packings may be quasiperiodic. Despite their relatively simple geometrical description, the true densest structures are then beyond the reach of our numerical algorithm because the algorithm relies on periodic boundary conditions. We thus consider an approach that can analyze the infinite-system size limit of these packings. This approach, however, rests on assumptions about the overall structure of packings that are not rigorously justified, and is found to be suboptimal in a few instances. 

\begin{figure}[h!]
\centering
\includegraphics[width=1.0\columnwidth]{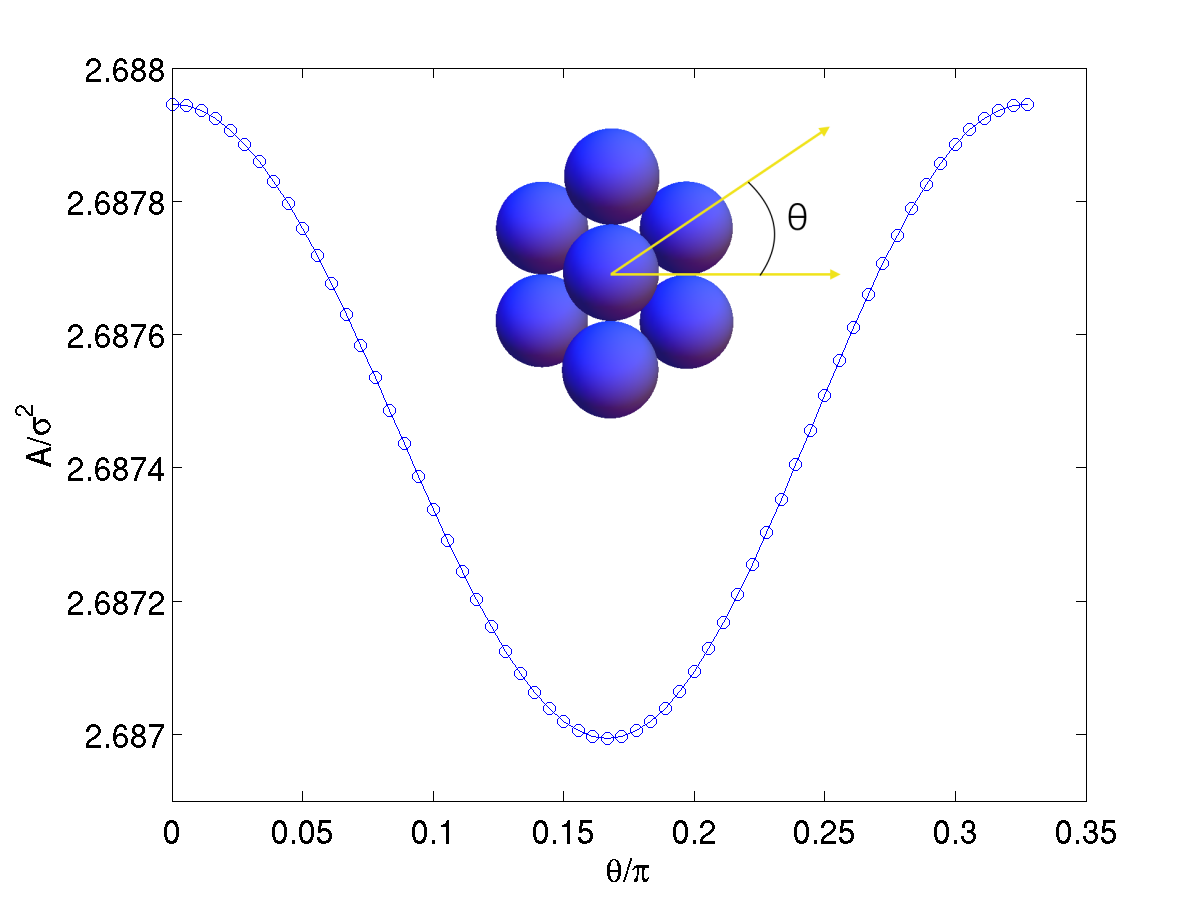}
\caption{The cylinder surface area occupancy for bent hexagons with different orientations for $D=3\sigma$. The area coverage is minimized for $\theta=\pi/6$ and is maximal for $\theta=0$, even though it is the latter structure that gives rise to the (6,6,0) outer shell.}
\label{fig:local}
\end{figure}

\begin{figure}[h!]
\centering
\includegraphics[width=1.0\columnwidth]{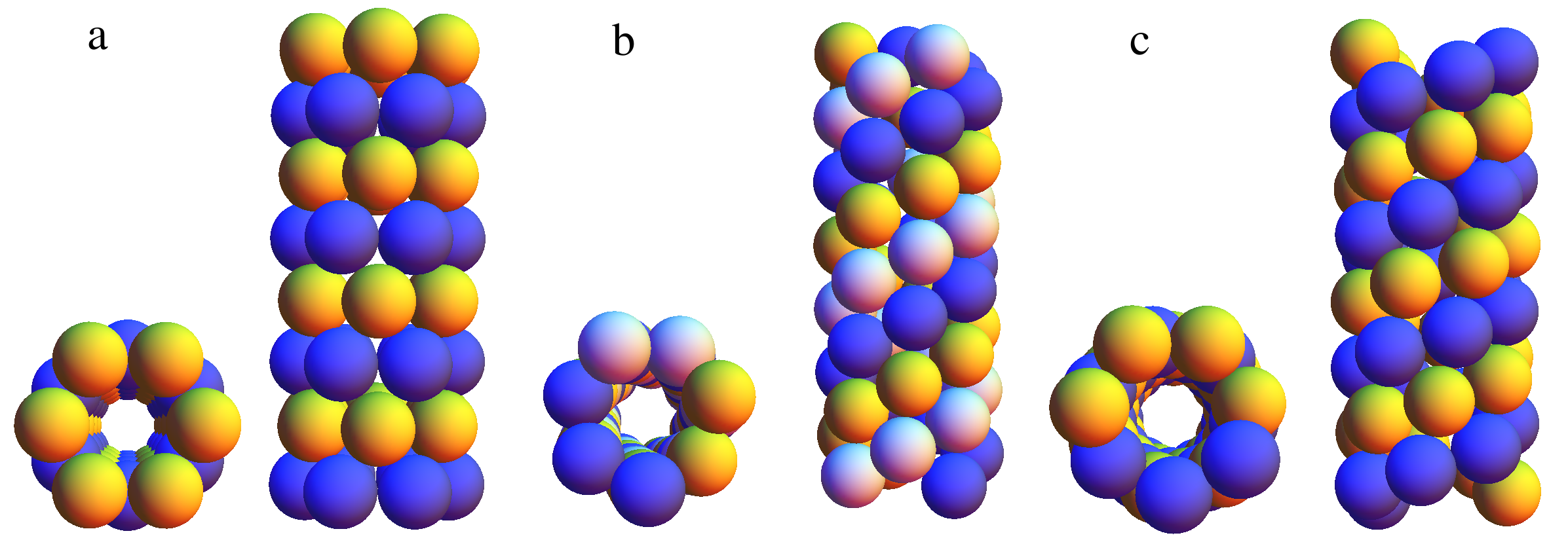}
\caption{Close-packed outer shells:  (a) stacked layers of 6 spheres each (6,6,0), (b) a 7-particle triple helix (7,4,3), and (c) a 7-particle double helix (7,5,2).}
\label{fig:shells}
\end{figure}

We first assume that packings in this regime have an outer shell that is as dense as possible along the cylinder wall.  For certain special values of $D$, these shells are close-packed, meaning that every sphere in that shell is in direct contact with six other particles within the shell (see Fig.~\ref{fig:shells}).  For $D=3\sigma$, for instance, the densest outer shell is the (6,6,0) structure.  Note, however, that even this seemingly straightforward case would be mathematically nontrivial to demonstrate. The structure cannot be argued to be optimal based on the local packing density, because a closed hexagon (or triangle) of spheres on the cylinder surface minimizes the covered area when one diameter of the hexagon is vertical rather than horizontal as in the (6,6,0) structure (Fig.~\ref{fig:local}). Proving that (6,6,0) shell is the densest possible for $D=3\sigma$ thus remains an open problem.  We nonetheless persist in this direction and next assume that each successive core particle then falls to its lowest possible position without perturbing the outer shell.  These assumptions are broadly consistent with the SLP results, and should allow for slightly denser structures to be obtained by side-stepping the periodicity constraint.

For convenience in the following, we define the shell density, $\rho_s$, as the number of shell spheres per unit length along the cylinder axis. As the cylinder diameter $D$ expands from a close-packed shell, the shell could expand radially and compress axially, but it is always preferable for a line slip to emerge instead~\cite{mughal2014pre}. It is an exercise in geometry to  analytically generate these structures. 
The density of the inner core is then computed as follows. We first choose a random height for the first particle and maximize its radial coordinate. Given the position of a sphere in the core, we assume the best way to pack the next higher sphere is to place it at the lowest possible position without moving any sphere in the shell or already placed in the core.  Each core sphere thus touches the core sphere below it and two spheres of the outer shell.  The procedure is iterated until the density of the core can be determined to within the desired accuracy.

In the special case of $D=3\sigma$, the procedure can be described by a map.  Recall that $\Delta_\perp = \sigma\sqrt{\sqrt{3}-1}$ is the spacing between successive layers of the shell.  Consider a core sphere at a generic specified height $z_1$ that is moved as far as possible off of the cylinder axis and is therefore touching two shell spheres.  Let $\xv_1$ be the position of this sphere and $\xv_2$ be the position of the core sphere that sits just above it.  Let $a_i = \{z_i/\Delta_\perp\}$, where $\{\cdot\}$ denotes the fractional part, indicate the relative height of the $i^{\mathrm{th}}$ sphere with respect to the shell layers just below and just above it.  We construct the map $M(a)$ that relates $a_2$ to $a_1$.  This map can then be iterated to determine the locations of all of the spheres in the core:
\begin{equation}
z_{i+1} = z_{i} + (1 + \{a_2 - a_1\} )\Delta_\perp\,. 
\end{equation}
If the map converges to a fixed point or a limit cycle, the core is periodic.  We will see, however, that this map is either quasiperiodic or has an extremely long period.

\begin{figure}[h!]
\centering
\includegraphics[width=1.0\columnwidth]{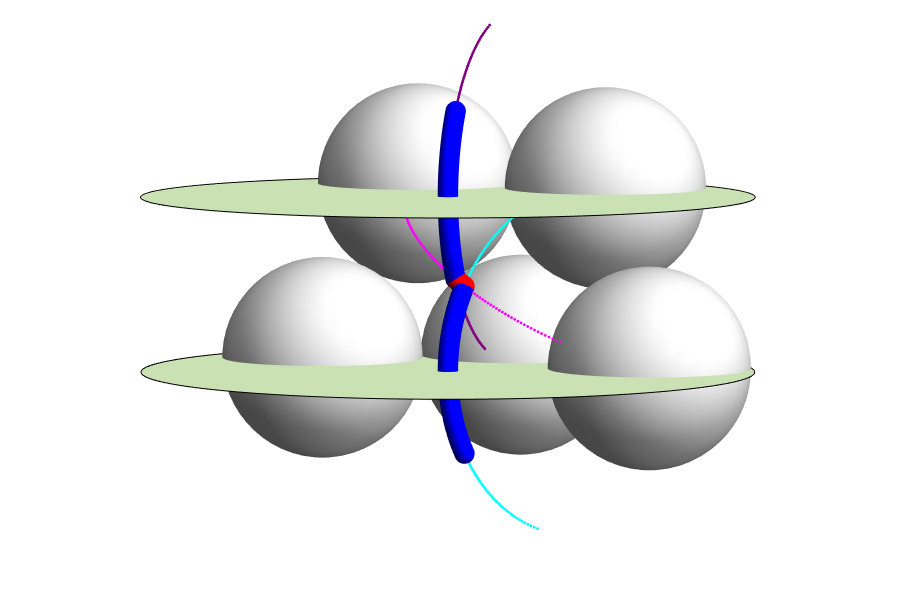}
\caption{Possible locations of a core sphere within a layer.  The large disks are cross sections of the confining cylinder, and are separated by $\Delta_\perp$.  The core sphere must be centered on a point on one of the thick blue curves or the short, thick red curve. See text for details.}
\label{fig:rolling6}
\end{figure}

Figure~\ref{fig:rolling6} shows the possible locations $\xv_1$ for a sphere within one layer.  The cyan circular arc shows the possible locations for a sphere that touches two adjacent shell spheres in a same layer.  The magenta arc shows the possible locations for a sphere that touches two adjacent spheres in different layers.  The purple arc shows the same locations as the cyan arc, but shifted up one layer and rotated accordingly by $\pi/6$ about the cylinder axis.  The portions of these arcs shown with thick blue and red curves (along with the positions related by the hexagonal rotations and reflections) are the possible locations of a core sphere within the depicted layer.  

\begin{figure}[h!]
\centering
\includegraphics[width=1.0\columnwidth]{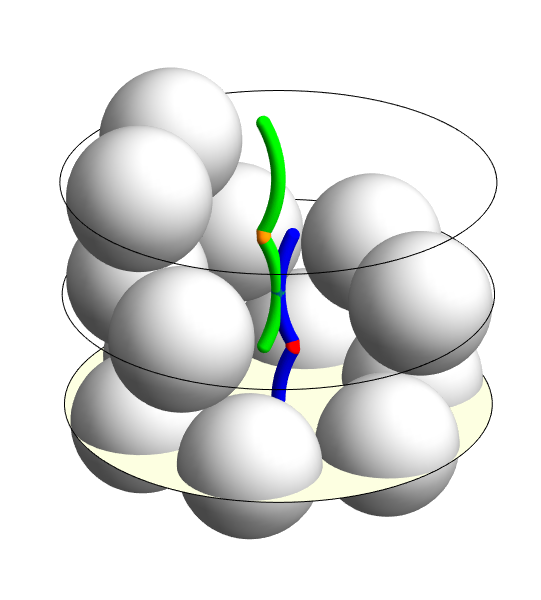}
\caption{Placement of a core sphere.  A point $P$ on the blue or red curve indicates a possible placement of a sphere center.  The sphere above it lies on the point on the green or orange curve that is a distance $\sigma$ from $P$. See text for details.}
\label{fig:mapcurves}
\end{figure}

The placement, $\xv_2$, of the sphere that rests on top of the sphere at $\xv_1$ must lie on a piecewise arc of the type shown in Fig.~\ref{fig:rolling6}, displaced one or two layers upward, and rotated so as to be as close to diametrically opposite $\xv_1$ as possible.  Inspection of the possible rotations and reflections reveals that the choice leading to the densest structure is a reflection through the plane containing the cyan arc in Fig.~\ref{fig:rolling6}, followed by a rotation by $7 \pi /6$ about the cylinder axis and a translation upward by $\Delta_\perp$, as shown in Fig.~\ref{fig:mapcurves}.

\begin{figure}[h!]
\centering
\includegraphics[width=1.0\columnwidth]{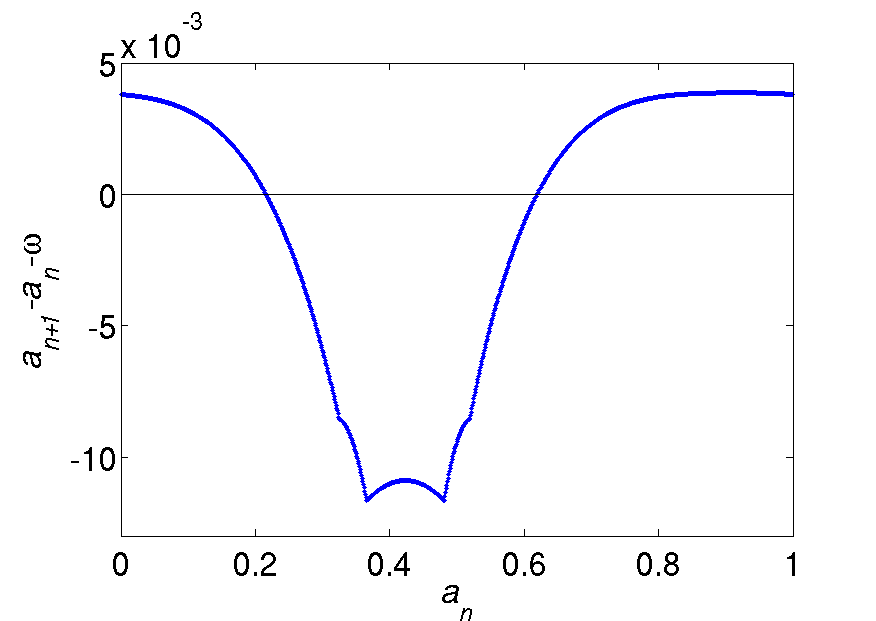}
\caption{Plot of the nonlinear term in the map from the height of one core sphere to the next one up, $f(a)$.  Note that for the purposes of this figure we have not taken the fractional part of $a_{n+1}$.  Note also the scale on the vertical axis.}
\label{fig:map6}
\end{figure}

The map $M(a)$ is determined by finding the value of $a_2$ along the upper curve (green/orange in Fig.~\ref{fig:mapcurves}) that is exactly $\sigma$ away from $a_1$ on the lower curve (blue/red).  $M(a)$ has the form of a circle map:
\begin{equation}
M(a) = \left\{ a + \omega + f(a)\right\}\,,
\end{equation} 
where $\omega$ is a constant and $f(a)$ is a periodic function with unit period and zero mean.  Numerical computation of the map yields $\omega = 0.163887$ and the function $f(a)$ shown in Fig.~\ref{fig:map6}.  Note the scale on the vertical axis; deviations from a line with unit slope are quite small.  The slope of $M(a)$ lies within the range $(0.8689,1.1533)$ everywhere, hence the map is monotonic and thus invertible, which means that it cannot be chaotic~\cite{Ot2002}. 

\begin{figure}[h!]
\centering
\includegraphics[width=1.0\columnwidth]{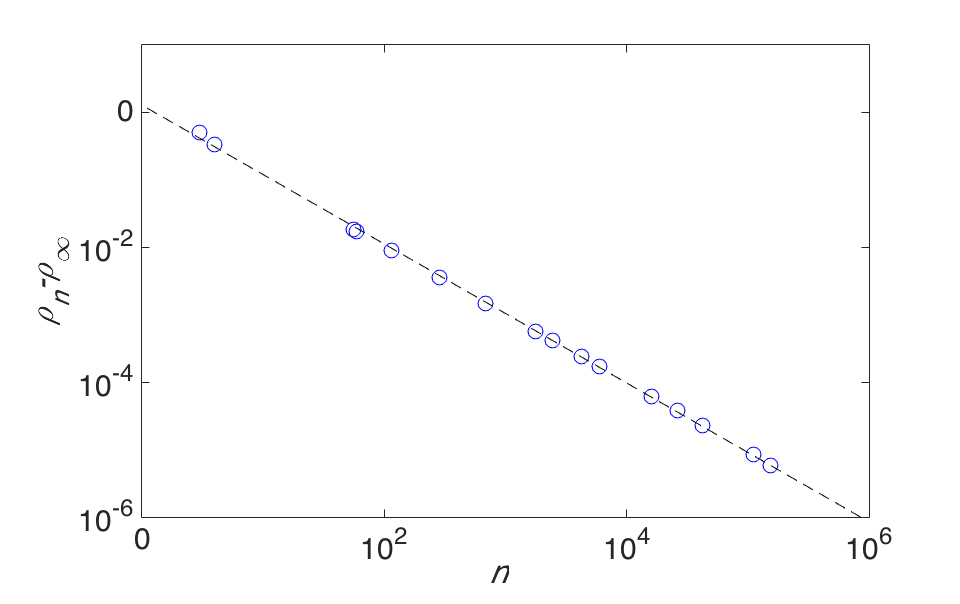}
\caption{Infinite-system size extrapolation of the packing density to $\rho_{\infty}$ for $D=3\sigma$ from the sinking algorithm.}
\label{fig:convergence}
\end{figure}

For a circle map with a small amplitude nonlinearity, the generic behavior is quasiperiodic~\cite{Ot2002}.  In the present case, we confirm the quasiperiodicity to a high degree of accuracy, i.e., we detect no exponential convergence to a limit cycle, and we determine the asymptotic density $\rho_{\infty}$ by fitting the values of $\rho_n = n/z_n$ to the form $\rho_{\infty} + c/n$, where $c$ is a constant. More precisely, we compute $z_n$, the height of the $n^{\mathrm{th}}$ sphere in the core, taking $z = 0$ to be a point at the center of a layer of the outer shell and beginning with $z_1 = 0$.  We then extract the sequence of $z_n$ values for which the $n^{\mathrm{th}}$ sphere sets a new record for coming closest to lying exactly in the plane of a layer, but is just below that plane, i.e, points for which $a_n$ comes ever closer to unity.  Note that if the sequence were converging to a periodic limit cycle, this procedure would either yield a finite sequence or one that exponentially approaches a value different from unity.  Figure~\ref{fig:convergence} shows $\rho_n - \rho_{\infty}$ as a function of $n$ for the sequence of best approximants, where $\rho_{\infty}$ has been adjusted to get a straight line on the log-log plot.  We find the best fit to be obtained for $\rho_{\infty} \approx 1.0043324(1)/\sigma$, which is slightly denser than for a simple stack of spheres on the cylinder axis, $1/\sigma$.
The total number of spheres per unit of cylinder length is thus $\rho_s + \rho_{\infty} =8.0169578(1)/\sigma$, corresponding to $\eta=0.593849(1)$, which is denser than $\eta=0.593661(1)$ obtained from the SLP algorithm as expected.

\begin{figure}[h!]
\centering
\includegraphics[width=1.0\columnwidth]{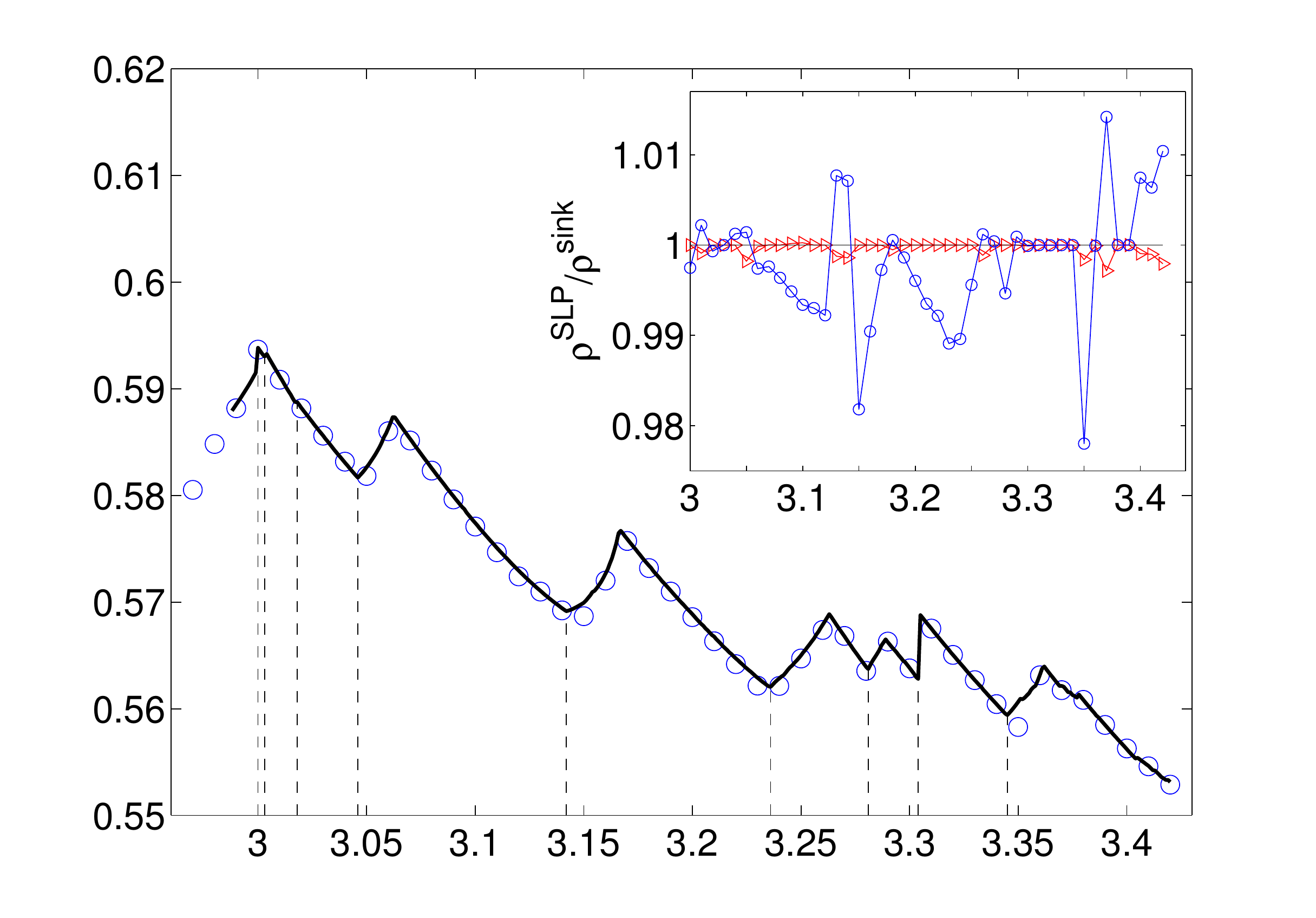}
\caption{Packing results for the SLP (blue circles) and the sinking (solid line) algorithms are in fairly close agreement, but the quasiperiodic phases identified by the latter are typically denser. The biggest differences are in the choice of optimal shell morphology around $D=3.01\sigma$ and $D=3.35\sigma$. The phase sequence for the sinking algorithm is (7,4,3), (6,6,0), (7,4,3), (6,6,0), (7,5,2), (7,6,1), (8,4,4), (8,5,3), (7,7,0) and (8,6,2), from left to right (separated by dashed lines). The inset shows the shell $\rho_{\mathrm{s}}^{\mathrm{SLP}}/\rho_{\mathrm{s}}^{\mathrm{sink}}$ (red triangles) and core $\rho_{\infty}^{\mathrm{SLP}}/\rho_{\infty}^{\mathrm{sink}}$ (blue circles) linear density ratio as a function of $D$. See text for details.}
\label{fig:sink}
\end{figure}

For $D\neq3\sigma$, the lack of reflection symmetry in the helical shells and the presence of line slips complicate the construction of a map from one core sphere height to the next.  We instead perform brute force numerical computations of the core packing algorithm for $n$ spheres and take the core density to be $\rho = n/z_n$.  The joint core and shell packing fraction, $\eta(D)$, is  obtained with a resolution of $\Delta D=0.001\sigma$ (see Fig.~\ref{fig:sink}). For most values of $D$, the sinking algorithm gives structures with a higher packing fraction than the SLP algorithm, but differences that are typically less than 0.1\%, which suggests that the SLP algorithm performs remarkably well in this regime. The most significant structural difference between the two algorithms is observed for $3.003\sigma\le D\le3.017\sigma$, where the sinking algorithm identifies the densest structures to have a (7,4,3) outer shell, while the SLP algorithm produces a (6,6,0) outer shell. The packing fraction difference between these two structures is small, but well above our numerical precision. The discrepancy may thus result from the former structure not being as easily accessible in the SLP search than the latter for our choice of algorithmic parameters and initial conditions.

At the level of precision considered for the SLP study ($\Delta D=0.01\sigma$), the sinking algorithm is found not to win outright at three points: $D=3.04\sigma$, $D=3.27\sigma$, and $D=3.40\sigma$ (see Table~\ref{tbl:diff}). We find two distinct mechanisms are at play here, which can be seen through consideration of the linear density ratio between the SLP and the sinking algorithms separately for the shell, $\rho_{\mathrm{s}}$, and for the core, $\rho_{\infty}$ (Fig.~\ref{fig:sink} inset). Interestingly, we find that $\rho_{\mathrm{s}}^{\mathrm{SLP}}/\rho_{\mathrm{s}}^{\mathrm{sink}}\le1$ 
at $3.04\sigma$ and $3.40\sigma$.
This indicates that a denser core is obtained at the expense of having a slightly perturbed (and less dense) outer shell.
At $3.27\sigma$, some shell spheres instead do not touch the cylinder wall (taking over some of the empty core space), which increases both the shell and the core densities.
(For $D=3.09-3.10\sigma$, the SLP also identifies a denser outer shell than the one used in the sinking algorithm, but in these cases the denser core obtained by the sinking algorithm more than compensates for this difference.)
The effect of the coupling between the outer shell and the inner core may be due to the proximity of a change to the shell symmetry ($D\approx3.04\sigma$), or to the end of this regime ($D\approx3.40\sigma$), but finer resolution studies would be needed to make more definitive statements.
It is also unclear whether this coupling is strong enough to make the densest structures periodic. SLP identifies unit cells that contain at least $55\le N\le80$ spheres, but more would be certainly possible. 

\begin{table}[h]
\small
  \caption{\ Density differences between SLP and sinking structures for points where the former is denser than the latter.}
  \label{tbl:diff}
  \begin{tabular}{@{\extracolsep{\fill}}c|c|c|c}
    \hline
    $D/\sigma$ & $(\rho_{\mathrm{s}}^{\mathrm{SLP}}-\rho_{\mathrm{s}}^{\mathrm{sink}})\sigma$ & $(\rho_{\mathrm{\infty}}^{\mathrm{SLP}}-\rho_{\mathrm{\infty}}^{\mathrm{sink}})\sigma$ & $\eta^{\mathrm{SLP}}-\eta^{\mathrm{sink}}$\\
    \hline
    3.04 & $-6.4\times10^{-6}$ & $1.259\times10^{-3}$ & $9.0\times10^{-5}$\\
    3.27 & $2.9\times10^{-6}$ & $4.510\times10^{-4}$ & $2.8\times10^{-5}$\\
    3.40 & $-7.9\times10^{-3}$ & $9.254\times10^{-3}$ & $7.2\times10^{-5}$\\
    \hline
  \end{tabular}
\end{table}

\section{Conclusion}

In this study, we have extended the range of known HS packings in cylinders of diameters $D=2.862\sigma$ to $D=4.00\sigma$ by adapting the SLP method of Ref.~\cite{torquato2010PRE} to this geometry and by developing a sinking algorithm. We have identified 17 new structures, most of them chiral, along with their continuous deformation. We also distinguish ranges of cylinder diameters over which different types of packings are observed. Most notably, around $D=3\sigma$ the outer shell is both fairly independent of the inner core and closely packed, and both algorithms provide strong numerical evidence that many of the packings are quasiperiodic. 

Although our study ended at $D=4.00\sigma$, we expect the competition between different shells to persist even once three and four of them develop. For $D\gg4\sigma$, however, the bulk FCC limit should eventually be recovered. The shell area should thus eventually form but a thin wrapping of a FCC core. Based on the analogy with packing of disks within a circle~\cite{hopkins2010pre}, however, we don't expect this phenomenon to develop before $D\gtrsim 20\sigma$, which is far beyond the regime that can be reliably studied with existing numerical methods. 

In closing, it is important to recall the difference between packings and their assembly from local algorithms. Although optimal, some of the packings may be dynamically hard to access (or even inaccessible) via self-assembly, which can be important in simulations and colloidal experiments. This aspect will be the object of a future publication.

\section*{Acknowledgments}
This work was supported by the National Science Foundation's Research Triangle Materials Research Science and Engineering Center (MRSEC) under Grant No. (DMR-1121107), and NSF grant from the Nanomanufacturing Program (CMMI-1363483). We thank Ce Bian, Daniela Cruz, Gabriel Lopez, R\'emy Mosseri, Crystal Owens, Wyatt Shields and Benjamin Wiley for stimulating discussions.

\bibliography{Cylinder.bib}
\appendix
\section*{Supplementary information}

\section{Unit cell sizes for packings identified to be optimal by the sequential linear programming (SLP) method}
SLP identifies the densest structures in the ranges $2.86\sigma\le D\le2.98\sigma$ and $3.43\sigma\le D\le4.00\sigma$ as well as for $D=3.04\sigma$, $3.27\sigma$ and $3.40\sigma$. 
\setlength\LTleft{0pt}
\setlength\LTright{0pt}
\begin{table*}[h]
\setlength\LTleft{0pt}
\setlength\LTright{0pt}
\small
  \caption{\ Structural parameters for different cylinder diameters. Quantities are rounded to the last digit. * denotes that the identified structure has a small enough period to give confidence that it is the truly optimal packing.  Other entries indicate the densest structure for the $N$ range we search. Recall that $\eta=[\frac{4}{3}\pi(\frac{\sigma}{2})^3N]/[\pi(\frac{D}{2})^2]\lambda_{\mathrm{z}}$.}
  \label{tbl:outer}
  \begin{tabular*}{\textwidth}{@{\extracolsep{\fill}}||c|c|c|c||c|c|c|c||c|c|c|c||}
    \hline
    $D/\sigma$ & $N$ & $\lambda_{\mathrm{z}}/\sigma$ & $\lambda_{\mathrm{\theta}}$ & $D/\sigma$ & $N$ & $\lambda_{\mathrm{z}}/\sigma$ & $\lambda_{\mathrm{\theta}}$ & $D/\sigma$ & $N$ & $\lambda_{\mathrm{z}}/\sigma$ & $\lambda_{\mathrm{\theta}}$\\
    \hline
    
    2.86 & 63 & 9.0393 & 2.9033 & 3.52 & 77 & 7.4307 & 0.6253 & 3.77 & 58 & 4.5814 & 0.8325\\
    2.87 & 50 & 6.9948 & 2.8225 & 3.53 & 77 & 7.3873 & 5.6060 & 3.78 & 58 & 4.5496 & 0.8230\\
    2.88 & 65 & 8.9876 & 5.2646 & 3.54 & 72 & 6.8727 & 3.7098 & 3.79 & 58 & 4.5220 & 0.8134\\
    2.89 & 138 & 18.9765 & 0.1512 & 3.55 & 62 & 5.8876 & 2.8030 & 3.80 & 54 & 4.1833 & 1.3869 \\
    2.90 & 117 & 15.9913 & 0.8482 & 3.56 & 62 & 5.8565 & 3.4986 & 3.81 & 54 & 4.1563 & 1.4064 \\
    2.91 & 81 & 10.9944 & 2.6513 & 3.57 & 52 & 4.8802 & 3.2919 & 3.82 & 54 & 4.1322 & 1.4034 \\
    2.92 & 141 & 18.9795 & 1.8146 & 3.58 & 66 & 6.0746 & 0.0044 & 3.83 & 54 & 4.1166 & 4.8716 \\
    2.93 & 90 & 12.0053 & 0.3859 & 3.59 & 66 & 6.0000 & 0.0153 & 3.84 & 54 & 4.1063 & 1.3987 \\
    2.94 & 83 & 10.9975 & 0.9297 & 3.60 & 55 & 4.9973 & 0.0563 & 3.85 & 79 & 5.9917 & 0.0480 \\
    2.95 & 69 & 1.9603 & 0.1637 & 3.61 & 66 & 5.9896 & 0.1373 & 3.86 & 79 & 5.9805 & 0.0473 \\
    2.96 & 69 & 9.0133 & 4.3213 & 3.62 & 72 & 6.5186 & 0.3233 & 3.87 & 71 & 5.3601 & 1.8866 \\
    2.97 & 69 & 8.9801 & 1.7785 & 3.63 & 57 & 5.1245 & 3.0348 & 3.88 & 71 & 5.3448 & 1.9012 \\
    2.98 & 55 & 7.0602 & 2.4692 & 3.64 & 83 & 7.4014 & 0.0912 & 3.89 & 71 & 5.3314 & 1.9129 \\
    3.04 & 48 & 5.9376 & 3.6792 & 3.65 & 83 & 7.3474 & 6.2378 & 3.90* & 12 & 0.8944 & 0.3191 \\
    3.27 & 50 & 5.4996 & 5.5196 & 3.66* & 12 & 1.0527 & 0.1915 & 3.91* & 12 & 0.8836 & 0.7146 \\
    3.40 & 78 & 8.0863 & 2.5233 & 3.67* & 12 & 1.0396 & 0.1961 & 3.92* & 12 & 0.8704 & 0.7083 \\
    3.43* & 10 & 1.0192 & 0.0000 & 3.68* & 12 & 1.0263 & 0.2005 & 3.93* & 12 & 0.8609 & 0.9076 \\
    3.44* & 10 & 1.0069 & 0.0000 & 3.69* & 12 & 1.0127 & 0.2050 & 3.94* & 12 & 0.8598 & 0.3492 \\
    3.45* & 10 & 1.0000 & 0.0097 & 3.70* & 12 & 0.9989 & 0.2094 & 3.95* & 14 & 1.0002 & 0.1737 \\
    3.46 & 59 & 5.9567 & 3.3920 & 3.71* & 12 & 0.9972 & 0.1993 & 3.96* & 14 & 0.9927 & 0.1797 \\
    3.47 & 65 & 6.4690 & 1.1549 & 3.72* & 12 & 0.9963 & 0.1953 & 3.97* & 14 & 0.9890 & 0.1822 \\
    3.48 & 51 & 5.0314 & 2.1022 & 3.73* & 12 & 0.9944 & 0.8389 & 3.98* & 14 & 0.9859 & 1.3955 \\
    3.49 & 51 & 4.9989 & 4.1921 & 3.74 & 74 & 6.0863 & 2.1105 & 3.99* & 14 & 0.9842 & 0.1769 \\
    3.50 & 51 & 4.9740 & 2.1083 & 3.75 & 50 & 4.0720 & 2.8097 & 4.00* & 14 & 0.9826 & 0.8123 \\
    3.51 & 72 & 6.9849 & 2.6963 & 3.76 & 63 & 5.0371 & 3.8694 & & & &\\
    \hline
  \end{tabular*}
\end{table*}

\end{document}